\documentclass[american,amssymb,amsmath,superscriptaddress,aps,pre]{revtex4-1}
\usepackage[latin9]{inputenc}
\usepackage{array}
\usepackage{amsmath}
\usepackage{graphicx}

\makeatletter

\usepackage{dcolumn}
\usepackage{bm}
\usepackage{subfigure}

\makeatother

\usepackage{babel}
\begin{document}

\title{Stability threshold approach for complex dynamical systems}

\author{Vladimir V. Klinshov$^{1*}$, Vladimir I. Nekorkin$^{1}$, J\"{u}rgen
Kurths$^{1,2}$ }

\address{$^{1}$Institute of Applied Physics of the Russian Academy of Sciences,
46 Ul'yanov Street, 603950, Nizhny Novgorod, Russia}

\address{$^{2}$Potsdam Institute for Climate Impact Research, Telegraphenberg
A 31, 14473, Potsdam, Germany}

\address{$^*$e-mail:vladimir.klinshov@ipfran.ru}

\begin{abstract}
A new measure to characterize stability of complex dynamical systems
against large perturbation is suggested, the stability threshold (ST).
It quantifies the magnitude of the weakest perturbation capable to
disrupt the system and switch it to an undesired dynamical regime.
In the phase space, the stability threshold corresponds to the ``thinnest
site'' of the attraction basin and therefore indicates the most ``dangerous''
direction of perturbations. We introduce a computational algorithm
for quantification of the stability threshold and demonstrate that
the suggested approach is effective and provides important insights.
The generality of the obtained results defines their vast potential
for application in such fields as engineering, neuroscience, power
grids, Earth science and many others where robustness of complex systems
is studied.
\end{abstract}

\maketitle

\section*{Introduction}

Complex systems science is strongly based on linear stability analysis
considering small perturbations of dynamical systems. In a seminal
paper \cite{PC1998} this concept was extended even to the stability
of synchronization in complex networks leading to the efficient master
stability formalism. However, for various applications often the influence
of large perturbations is also of crucial importance. Typical examples
are climatological systems, in particular ocean circulations. Well
accepted is that the Atlantic Meridional Overturning Circulation may
be sensitive to changes in the freshwater balance of the northern
North Atlantic. When an anomalous freshwater flux is applied in the
subpolar North Atlantic, this circulation collapses in many ocean-climate
models \cite{MOC}. Another example is power grids which are networks
of connected generators and consumers of electrical power. For proper
function of such networks synchronization between all the nodes is
essential. Local failures, overloads or lines breaks may cause desynchronization
of nodes and lead to large-scale blackouts \cite{Ewart78,Machovski2008}.

The study of system's stability against large perturbations implies
treating the following challenging problem: definition of the class
of ``safe'', or admissible perturbations after which the system
returns back to the initial regime. In contrast, ``unsafe'' perturbations
switch the system to another, often unwanted, dynamical regime. The
definition of the class of safe perturbations of a nonlinear system
is very complicated and basically different from the linear stability
analysis. The reason is that for large perturbations linearization
is inadequate and the perturbed dynamics is governed by nonlinear
equations whose analytical study is impossible in general. Some analytical
methods do exist, for example the method of Lyapunov functions \cite{Lyap}.
However, this method has serious limitations since a Lyapunov function
for a particular dynamical system is often not constructive. The ``safe''
perturbation class can also be analytically estimated for some specific
systems, e.g. networks of spiking neurons \cite{Timme}. Nevertheless,
an important task is to develop numerical methods of defining and
describing the class of safe perturbations \cite{Wiley2006}. 

\begin{figure*}[t]
\includegraphics[width=100mm]{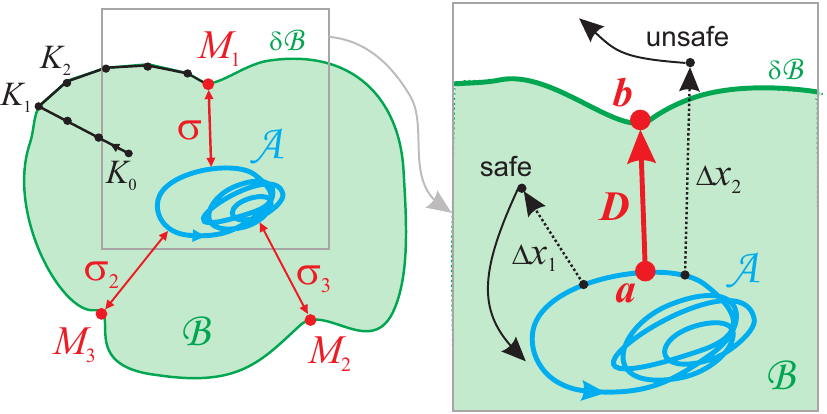}

\protect\caption{\label{fig:1} Stability threshold (ST) and its quantification. Attractor
$\mathcal{A}$, its attraction basin $\mathcal{B}$ and ST $\sigma$.
The trace of the algorithm converging to the point $M$ is shown by
black dots. Other LOCT points $M_{2}$ and $M_{3}$ are also shown.
In the the zoomed part, safe and unsafe perturbations are shown. Dotted
lines are perturbations, solid black lines are trajectories of the
perturbed system. }
\end{figure*}

From the viewpoint of nonlinear dynamics, established dynamical regimes
of the system corresponds to attractors in the phase space. The class
of safe perturbations is equal to the attraction basin, i.e. the
set of the points which converge to the attractor. A perturbation
is safe if it brings the system to a point within the attraction basin.
The first attempt to characterize attraction basins in complex networks
was undertaken in \cite{BS} where the concept of basin stability
 was introduced. The basin stability equals
\begin{equation}
S_{B}=\int_{Q}\chi(x)\rho(x)dx,\label{eq:SB}
\end{equation}
where $Q$ is the set of possible perturbed states $x$, $\rho(x)$
with $\int_{Q}\rho(x)dx=1$ is the density of the perturbed states,
and $\chi(x)$ equals one if the point $x$ converges to the attractor
and zero otherwise. The value $S_{B}\in(0;1]$ expresses the likelihood
that the perturbed system returns to the attractor. An important advantage
of this measure is that it can be easily calculated by Monte-Carlo
method. Namely, one should just take a large number of random points
in the phase space and check how many of them converge to the attractor.
If $M$ of $N\gg1$ points converge, $S_{B}\approx M/N$.

Basin stability is an important characteristic extending the concept
of linear stability for the case of large perturbations. However,
many real dynamical systems, especially complex networks, possess
highly-dimensional phase space with complicated structure. This makes
it problematic to characterize an attraction basin by just a single
scalar value. Moreover, basin stability depends on the perturbation
class $Q$ which should be chosen a priory. 

In this paper we suggest a new measure to characterize stability against
large perturbation, the \textbf{stability threshold} (ST). We were
inspired by the observation that for real systems it is often important
to know the maximal magnitude of perturbation which the system is
guaranteed to withstand, like the maximal voltage jump for a stabilizer
or the maximal bullet energy for a bulletproof vest. In the following,
in Sec. 1 we introduce the ST in detail and explain how to calculate
it. In Sec. 2 and 3 its potential is demonstrated for two paradigmatic
model systems. In Sec. 4 we relate the two measures, the stability
threshold and the basin stability. In Sec. 5 we briefly summarize
and discuss our results.

\section{Definition and quantification of stability threshold}

We define ST as the minimal magnitude of a perturbation capable to
disrupt the established dynamical regime, i.e. to push the system
out of the attraction basin. In the phase space, ST is the minimal
distance between the attractor $\mathcal{A}$ and the border $\mathcal{\delta B}$
of its attraction basin, i.e. 

\begin{equation}
\sigma=\inf\left\{ \mbox{dist}(a,b)\,|a\in\mathcal{A},\, b\in\mathcal{\delta B}\right\} ,\label{eq:sigma}
\end{equation}

where $\mbox{dist}(\cdot,\cdot)$ is the distance between the points
in the phase space. Further, we use Euclidean metric to calculate
the distance. However, any other metrics can be used as well. 

To better understand the physical meaning of ST consider the system
settled to the attractor $\mathcal{A}$ as depicted in Fig. 1. Let
$\, a\in\mathcal{A}$ and $b\in\mathcal{\delta B}$ be points corresponding
to ST such that $\mbox{dist}\left(a,b\right)=\sigma$. Consider now
a perturbation of the system which results in a quick change of its
state. Namely, let the system state change from the unperturbed one
$x_{u}$ to the perturbed one $x_{p}$ so that the perturbation $\Delta x=x_{p}-x_{u}$
has the magnitude $q=|\Delta x|$ . If $q<\sigma$, the perturbation
can never kick the system out of the attraction basin ($\Delta x_{1}$
in Fig. 1). But if $q>\sigma$ and the system is near the point $a$
just before the perturbation, it may be kicked out of the basin if
the direction of the vector $\Delta x$ is close to the vector $D=b-a$
($\Delta x_{2}$ in Fig. 1). The above reasoning shows that besides
the value of $\sigma$, the direction of the corresponding vector
$D$ is critical. This vector corresponds to the most ``dangerous''
direction of perturbations in which the distance to the basin border
is the shortest. 

Let us now come to a question of quantification of the ST. For this
purpose we suggest a two-stage algorithm whose basic principles are
described below and also illustrated in Fig. 1. 

i) On the first stage we identify some point $K_{1}$ on the border
of the attraction basin. For this purpose we choose an arbitrary point
$K_{0}$ in the vicinity of the attractor and start to move from the
attractor until leaving the basin. The point $K{}_{1}$ is found then
by the bisection method. 

ii) On the second stage we move along the basin border. On each step
we draw a tangential hyperplane to the border at the current point
$K_{n}$. In the hyperplane we find the point closest to the attractor
$\mathcal{A}$ and make a step towards this point and so obtain the
new point $K_{n+1}$. Such steps bring us closer and closer to the
attractor and finally converge to the point $M$ on the border with
the minimal distance to the attractor . 

The second stage relies on smoothness of the border for only in this
case it can be approximated by a tangential hyperplane. Further we
assume the border is smooth (defined by a function of class $C^{1}$)
which is the case for many realistic systems.

The suggested algorithm allows us to determine the local minima of
the distance between the attractor and the basin border, which we
call further ``local threshold'' (LOCT) points. Starting from different
initial points we get different LOCT points $M_{1},M_{2},...,M_{m}$
(Fig. 1). Between them, the one closest to the attractor is the ``global
threshold point'' corresponding to the ST: $\sigma=\min(\sigma_{1},...,\sigma_{m}),\textrm{\;\mbox{ where }}\sigma_{j}=\mbox{dist}\left(M_{j},\mathcal{A}\right).$ 

This brute-force search to obtain all the local minima does not seem
to be a very effective strategy. However, effectiveness of the method
is essentially improved in a parametric study, i.e. when the system
properties are studied versus its parameters. Note that such tasks
are typical since all realistic systems depend on parameters and usually
one wants to know what happens if they are varied. Suppose that for
a certain parameter value $p=p_{0}$ we have found all LOCT points
$M_{1}(p_{0}),...,M_{m}(p_{0})$. In a robust system, the phase space
structure changes continuously when $p$ is changed. Thus, the coordinates
of LOCT points depend continuously on $p$. So, when $p$ is changed
by a small value $\Delta p=p-p_{0}$ one should start the algorithm
from the points $M_{j}(p_{0})$. Since the actual positions of $M_{j}(p)$
are close, the algorithm converges to them quickly. In this manner
one can effectively trace the positions of LOCT points over the parameter
value.

One can still argue that to find all the LOCT points even for one
parameter value $p_{0}$ may be a practically impossible task for
complex high-dimension systems. However, it is important to emphasize
that this task is significantly simplified if the system is a network,
i.e. it is composed of many low-dimensional subsystems. In this case
the coupling strength is a natural choice for the parameter $p$ one
can trace the system over. For the case of no coupling ($p=0$) the
high-dimension phase space of the whole system is just a direct product
of the low-dimension subspaces. In each of these subspaces the LOCT
points can be found relatively easily which allows to spot them in
the full space as well. Further one can gradually increase the coupling
and trace the points.

Another issue is the possible emergence of new LOCT points. We have
an effective method to trace once found points over the parameter,
but how can one assure that no other points have emerged closer to
the attractor? This problem is typical for the global extremum seeking
in the case when full sampling of the space is impossible \cite{Strongin2000}.
Although there is no way to guarantee that the found ST point is indeed
the global minimum, there is a way to make the probability of that
arbitrarily close to one. For this sake one has to check sufficiently
many random initial points and make sure that none of them gives a
better result. We revisit this issue in the end of Sec. 4 where we
provide an estimate of the number of trials one should run to decrease
the mistake probability below a given level.

\section{Stability threshold of pendulum}

Now we show how the ST approach can be applied to study some paradigmatic
dynamical systems. First we consider a classic pendulum under an external
force $P$:

\begin{equation}
\frac{d\theta}{dt}=\omega,\qquad\frac{d\omega}{dt}=-\alpha\omega+P-\sin\theta.\label{eq:pendulum}
\end{equation}

Here, $\theta$ and $\omega$ are the deviation angle and the angular
velocity, and $\alpha$ describes friction. Noteworthy models similar
to (\ref{eq:pendulum}) are often used to describe dynamics of nodes
of power grids, i.e. generators or consumers \cite{Machovski2008,Rohden2012}.
The phase space of the model is a cylinder $S^{1}\times R^{1}$ and
includes two attractors: a stable steady state $O(\arcsin P,0)$ and
a stable limit cycle $L$ (Fig. 2a). In the context of power grids,
the steady state corresponds to the state when the generator operates
in synchrony with the grid, and the limit cycle corresponds to an
undesired asynchronous regime.

\begin{figure*}[t]
\includegraphics[width=100mm]{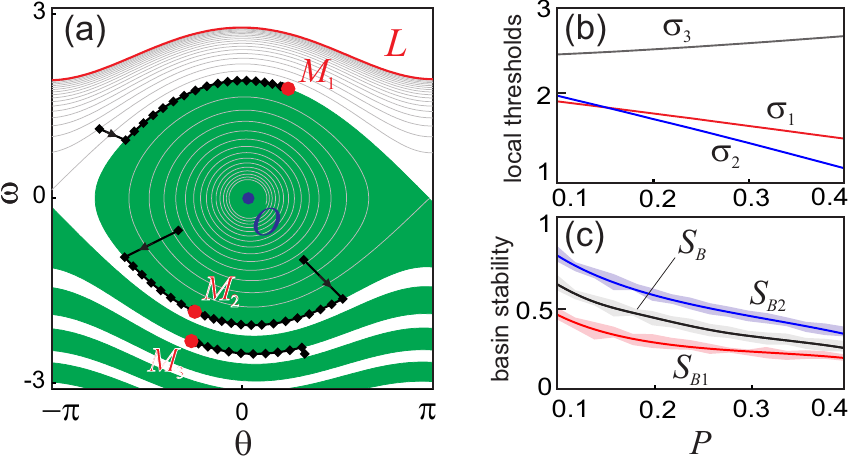}

\protect\caption{\label{fig:2}Stability of the pendulum (\ref{eq:pendulum}). (a)
Phase space for $\alpha=0.04$, $P=0.1.$ Green area is the attraction
basin of the steady state $O$, red curve is the limit cycle $L$.
LOCT points are depicted by red dots, traces of the algorithm by black.
(b) Local stability thresholds $\sigma_{1}$ (red), $\sigma_{2}$
(blue), and $\sigma_{3}$ (black) versus $P$. (c) The basin stability
values $S_{B1}$ (red), $S_{B2}$ (blue) and $S_{B0}$ (black) versus
$P$, mean values and variances. }
\end{figure*}

Next we use the concept of ST to study the attraction basin of the
steady state $O$. The identified LOCT points are depicted by red
dots in Fig. 2a. The most important ones are $M_{1}$ corresponding
to positive perturbations and $M_{2}$ corresponding to negative ones.
Because of the complex shape of the attraction basin other LOCT points
exist further from the attractor, e.g. $M_{3}.$ Figure 2b demonstrates
the local thresholds (LOCTs) $\sigma_{j}=\mbox{\mbox{dist}}(O,M_{j})$
associated with these points in dependence on the parameter $P.$
One can see that for $P<P^{*}\approx0.15$ the point closest to the
attractor is $M_{1},$ while for $P>P^{*}$ the closest point is $M_{2}$.
Thus, the ST equals $\sigma_{1}$ for $P<P^{*}$ and $\sigma_{2}$
for $P>P^{*}$, i.e. the most dangerous are positive perturbations
for small $P$ but negative perturbations for large $P$. 

It is interesting to compare both basin measures: stability threshold
and basin stability. For this sake $S_{B}$ is plotted versus $P$
in Fig. 2c. We calculate it for three different classes of perturbations:
positive perturbations ($S_{B1}$ for $Q_{1}=[-\pi;\pi]\times[0;3]$),
negative perturbations ($S_{B2}$ for $Q_{2}=[-\pi;\pi]\times[-3;0]$),
and perturbations of both signs ($S_{B0}$ for $Q=Q_{1}\cup Q_{2}$).
When $P$ increases, the basin stability for all classes of perturbations
decreases as well as the ST. Thus, both measures indicate that the
system becomes less robust. However, basin stability fails to detect
which perturbations are more dangerous: $S_{B2}$ is sufficiently
larger than $S_{B1}$ for all values of $P$.

We also checked that the efficiency of our algorithm is essentially
improved by tracing LOCT points over the parameter. For this sake
we identified the position of the point $M_{1}$ for $P\in[0.1;0.4]$
for three different setups: parameter step $\Delta P=0.02$ (16 datapoints),
without tracing; the same parameter step, with tracing; smaller step
$\Delta P=0.01$ (31 datapoints), with tracing. Without tracing, the
search started each time from the same point. With tracing, the search
for the new parameter value started from the position found for the
previous parameter value. The total computation time $T_{c}$ equals
$9\times10^{-3}$(a.u.) for the first setup, $19\times10^{-4}$ for
the second setup, and $24\times10^{-4}$ for the third setup. Thus,
with tracing the computation time decreases approximately five times.
For higher-dimensional systems the improvement is even much higher.
Note also that $T_{c}$ in the third setup increases by less than
30\% with respect to the second setup, although the number of datapoints
is twice larger. The reason is that with a smaller parameter step
the positions of the LOCT points change less and they are found faster.

\section{Stability threshold of complex networks}

The second example is a network of coupled one-dimensional maps. We
chose maps for two reasons: first, because of simpler implementation,
and second, to demonstrate the generality of our approach. The network
on $N$ nodes is governed as follows:

\begin{equation}
x_{i}(t+1)=ax_{i}(t)+bx_{i}^{2}(t)+\kappa\sum_{j=1}^{N}c_{ij}\left(x_{j}(t)-x_{i}(t)\right).\label{eq:sqnetwork}
\end{equation}

Here, $0<a<1$ is the system parameter, $\kappa$ stands for the global
coupling coefficient and $c_{ij}$ are the elements of the coupling
matrix. Coupling between two nodes $i$ and $j$ equals $\kappa c_{ij}$.
The network has the only attractor, the stable fixed point $O(0,0,...,0)$.
However, after a large perturbation the system trajectories may go
to infinity.

For network (\ref{eq:sqnetwork}), a natural way to find LOCT points
is to trace them over the coupling coefficient $\kappa$. For $\kappa=0$,
the nodes are uncoupled and each of them is governed by the map $x_{i}(t+1)=ax_{i}(t)+x_{i}^{2}(t)$,
which has a stable fixed point $x_{i}=0$ with the attraction basin
$-1<x_{i}<1-a$. The borders of this interval define two LOCT points
in the network phase space: $M_{i+}\left(x_{j}=0\,(j\neq i),\, x_{i}=1-a\right)$
corresponds to positive perturbation of the node $i$, and $M_{i-}\left(x_{j}=0\,(j\neq i),\, x_{i}=-1\right)$
to negative ones. We start from these points for $\kappa=0$, then
gradually increase $\kappa$ and trace their positions. We also periodically
check for emergence of new LOCT points, but failed to detect any.

\begin{figure*}[t]
\includegraphics[width=100mm]{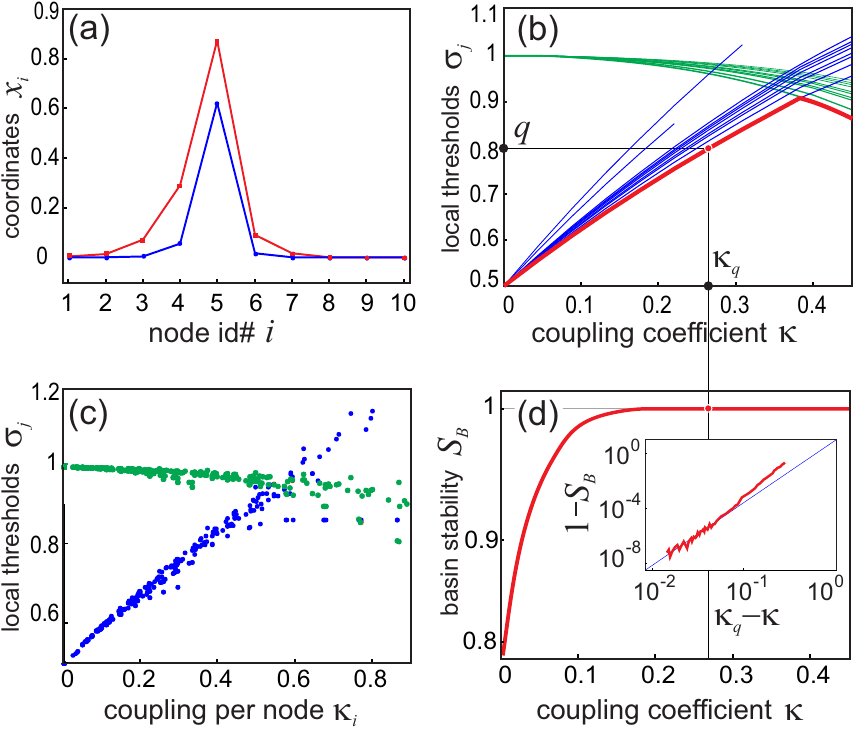}

\protect\caption{\label{fig:3}Stability of the network (\ref{eq:sqnetwork}). (a)
Coordinates of a typical LOCT point $M_{i+}$ ($i=5)$ for two values
of $\kappa$ - small (blue) and large (red). (b) LOCTs $\sigma_{i\pm}$
versus $\kappa$. Blue thin curves for $\sigma_{i+}$, green thin
curves for $\sigma_{i-}$. Red thick curve is the ST $\sigma$. (c)
LOCTs $\sigma_{i\pm}$ versus nodal coupling strength $\kappa_{i}$
for various nodes, coupling coefficients and network configurations.
Blue dots for $\sigma_{i+}$, green for $\sigma_{i-}$ (d) The basin
stability versus $\kappa$. The inset shows the same dependency in
logarithmic scale. Red curve for numerical results, blue thin line
for the estimate (\ref{eq:scaling}).}
\end{figure*}

We study various networks with $2\leq N\leq100$ and different types
of topology: all-to-all, random \cite{ER1960}, small-world \cite{Watts1998},
scale-free \cite{Barabasi1999}, and cluster networks \cite{Klinshov2014}.
In all the cases, the behavior of LOCT points is quite similar. When
$\kappa$ increases, the positions of the points change, so that the
coordinates $x_{j}$ ($j\neq i$) of $M_{i\pm}$ are no longer zeros.
However, for weak coupling the coordinates of LOCT points obey $|x_{i}|\gg x_{j}$,
i.e. the corresponding perturbation mainly concerns the node $i$.
For larger $\kappa$ the situation changes and LOCT points may have
several coordinates of the same order. Typical LOCT points are illustrated
in Fig. 3a. 

Now let us consider LOCTs $\sigma_{i\pm}=\mbox{dist}(O,M_{i\pm})$
associated with LOCT points. A typical dependence of these thresholds
on $\kappa$ is illustrated in Fig. 3b. For all $i$, $\sigma_{i+}$
grows with $\kappa$, while $\sigma_{i-}$ decreases. Some of the
points $M_{i+}$ may disappear at certain $\kappa$ as well. A detailed
study shows a remarkable feature of the LOCTs $\sigma_{i\pm}$: they
turn out to be strongly correlated with the values of total connections
strength to the node $\kappa_{i}=\kappa\sum_{j=1}^{N}c_{ij}$. In
Fig. 3c the LOCTs $\sigma_{i\pm}$ are plotted versus $\kappa_{i}$
for various nodes, coupling coefficients, network sizes and configurations.
The correlation is large, especially for small $\kappa_{i}$. Notice
that for $\kappa_{i}\lesssim\kappa^{*}\approx0.6$ positive perturbations
have a lower threshold than negative ones, and this threshold increases
with $\kappa_{i}.$ This finding leads to an easy and intuitively
clear rule: the stronger the node is connected to the network the
harder it is to tear it off. However, too strong coupling ($\kappa_{i}\gtrsim\kappa^{*}$)
is undesirable, since it increases susceptibility to negative perturbations. 

The global ST of the network is defined by the lowest LOCT. Figure
3b illustrates a typical dependence of the ST on $\kappa$.

\section{Stability threshold and basin stability}

It is interesting to compare the two measures of stability against
large perturbations, the stability threshold and the basin stability
for the same network (Fig. 3d). As the perturbation class $Q$ we
use a hypersphere of radius $q=0.8$ with constant density $\rho$
which means that we consider perturbations of amplitude $q$ and random
direction. \cite{Muller1959} One may see that $S_{B}=1$ when the
ST exceeds $q$ for $\kappa>\kappa_{q}\approx0.26$. This confirms
that the ST indeed characterizes the weakest perturbation that can
disrupt the network.

From Fig. 3d one may acquire the wrong impression that the basin stability
reaches unity much earlier than $\kappa$ reaches $\kappa_{q}$. The
reason is that $S_{B}$ approaches unity very quickly when $\sigma$
approaches $q$. This can be seen in the inset of Fig. 3d which has
a logarithmic scale. This feature seems to be typical for high-dimensional
dynamical systems. Indeed, consider an arbitrary dynamical system
in the $N$-dimensional phase space settled into the attractor $\mathcal{A}$
with the stability threshold $\sigma.$ Consider the perturbation
class $Q$ consisting of perturbations with the amplitude $q$. For
$q<\sigma$, the set $Q$ resides inside the attraction basin $\mathcal{B}$,
therefore $S_{B}=1$. For $\sigma=q$, the set $Q$ contacts the border
of the basin $\delta\mathcal{B}$. For $q>\sigma$ some part of the
set $Q$ gets out of the basin $\mathcal{B}$ and $S_{B}$ becomes
smaller than one (Fig. 4a). The probability of the perturbed state
to be out of the basin is proportional to the surface area $s$ of
the protrusive part (gray in the figure), so $1-S_{B}\sim s$. To
estimate the surface area, one can approximate both surfaces $Q$
and $\delta\mathcal{B}$ by quadratic forms near the site of their
intersection. Then, the transverse size of the protrusive part can
be estimated as $d\sim\sqrt{q-\sigma}$, and the surface area $s\sim d^{N-1}$.
This leads to the estimate 

\begin{equation}
1-S_{B}\sim(q-\sigma)^{\frac{N-1}{2}}.\label{eq:scaling}
\end{equation}
The corresponding slope is given by the blue line in the inset of
Fig. 4d and agrees with the numerical results. 

\begin{figure*}[t]
\includegraphics[width=100mm]{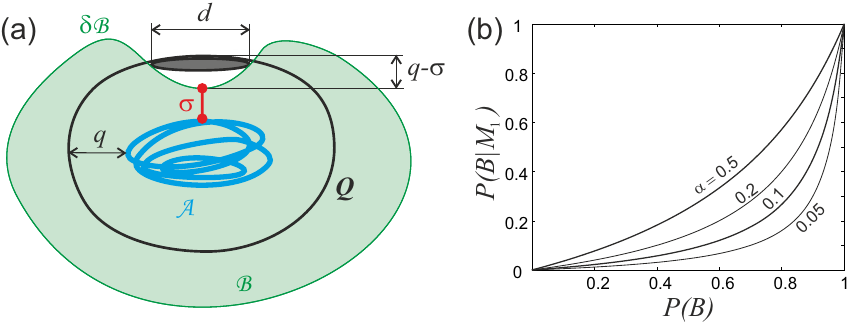}

\protect\caption{\label{fig:4}(a) Estimate of basin stability for perturbations slightly
exceeding the ST. (b) The posterior probability $P(B|M_{1})$ versus
the prior probability $P(B)$ for various values of $\alpha$ denoted
on the figure. }
\end{figure*}

The scaling law (\ref{eq:scaling}) suggests that for high-dimensional
systems it is very unlikely that the system will be disrupted by a
random perturbation of magnitude just above the ST. From the other
side, a wisely designed perturbation can disrupt the system even being
slightly above the ST. 

From computational prospective, the estimate (\ref{eq:scaling})
shows that attempts to estimate the stability threshold from basin
stability is inefficient for high-dimensional systems. Indeed, on
the example from Fig. 4d one can see that it is very complicated to
detect the exact point where the basin stability reaches exactly unity.
The probability to randomly hit a point which is $\varepsilon$ above
the stability threshold is of the order of $\varepsilon^{(N-1)/2}$,
so the time to determine the threshold with the accuracy $\varepsilon$
by the brute-force strategy grows inversely. In contrast, the method
suggested herein allows to reach the LOCT point and determine the
ST in quite a few steps.

The estimate (\ref{eq:scaling}) can be also used to determine the
probability of a wrong conclusion about the ST. Suppose we have a
potential ST, i.e. a global minimum point $M_{1}$ with $\sigma_{1}=\mbox{dist}\left(M_{1},\mathcal{A}\right)$
and want to check if there exists another LOCT point $M_{2}$ with
smaller $\sigma_{2}=\mbox{dist}\left(M_{2},\mathcal{A}\right)=\sigma_{1}-\delta\sigma$.
For this sake we choose random starting points $K_{1}$ at $q=\sigma_{1}+\varepsilon$
until finding a point out of the basin and run the algorithm. If it
converges to a point $M_{2}$ different from $M_{1}$ and with $\sigma_{2}<\sigma_{1}$
we have found a new potential ST. And if it converges to $M_{1}$
two competing hypotheses are to be evaluated:

Hypothesis $A$. The point $M_{1}$ indeed corresponds to the ST and
there is no other point $M_{2}.$ In this case, the algorithm converges
to $M_{1}$ with probability $P(M_{1}|A)=1$.

Hypothesis $B=\overline{A}$. The point $M_{2}$ exists, but we have
not found it. According to (\ref{eq:scaling}), the probabilities
to converge to $M_{1}$ and $M_{2}$ relate as

\begin{equation}
\frac{P(M_{1}|B)}{P(M_{2}|B)}=\alpha\sim\frac{(q-\sigma_{1})^{\frac{N-1}{2}}}{(q-\sigma_{2})^{\frac{N-1}{2}}}=\left(\frac{1}{1+\frac{\delta\sigma}{\varepsilon}}\right)^{\frac{N-1}{2}}.
\end{equation}

According to Bayes' theorem, the posterior probability $P(B|M_{1})$
is expressed through the prior probability $P(B)$ as

\begin{eqnarray}
P(B|M_{1}) & = & \frac{P(M_{1}|B)P(B)}{P(M_{1}|A)P(A)+P(M_{1}|B)P(B)}=\nonumber \\
 & = & \frac{1}{1+\left(\frac{1}{\alpha}+1\right)\left(\frac{1}{P(B)}-1\right)}.\label{eq:PBM1}
\end{eqnarray}

In Fig. 4b the posterior probability is plotted versus the prior probability
for several values of $\alpha$. Note that if $\alpha$ is small enough
and $P(B)$ is not close to one (\ref{eq:PBM1}) can be approximated
as $P(B|M_{1})\approx\alpha P(B)$, which means that each trial decreases
the probability by the factor $\alpha$. Thus, after $M$ trials the
posterior probability of $B$ can be estimated as

\begin{equation}
\ln P_{M}\sim-\frac{M(N-1)}{2}\ln\left(1+\frac{\delta\sigma}{\varepsilon}\right).\label{eq:lnPM}
\end{equation}
The estimate (\ref{eq:lnPM}) gives the probability that after $M$
trials we have still missed a LOCT point whose distance to the attractor
is by $\delta\sigma$ lower than $\sigma_{1}$. By sufficient increasing
of $M$ this mistake probability can be made arbitrarily small.

\section{Conclusions and discussion}

To conclude, we have introduced a novel measure to describe stability
of dynamical systems against external perturbations. This is the stability
threshold (ST) which equals the magnitude of the weakest perturbation
capable to disrupt the established dynamical regime. The ST provides
important information, since it guarantees the system to withstand
any perturbation of smaller magnitude. In the phase space, the ST
is the minimal distance between the system's attractor and the border
of its attraction basin. From this prospective, the ST defines the
``thinnest site'' of the basin. And as the saying goes, where something
is thin, that is where it tears: the direction corresponding to ST
is the most dangerous for the system.

For dynamical networks, different directions in the multidimensional
phase space are associated with different nodes. To this end, the
ST approach allows to determine the nodes which are mostly susceptible
to perturbations. Applying external perturbations to these nodes,
one may disrupt the network comparatively easily. However, sometimes
the ST is associated with perturbations involving several nodes. An
example of such a situation is depicted in Fig. 3a. Under such circumstances,
it is easier to disrupt the network by simultaneous perturbation of
several nodes rather than by perturbing just one of them. 

We have also suggested an algorithm to calculate the ST for arbitrary
dynamical systems and demonstrated its effectiveness. The generality
of the ST-based approach defines its vast potential for applications.
Possible fields include engineering, neuroscience, power grids, Earth
science and many others where robustness of complex systems against
large perturbations is important.

Besides application to the study of particular systems, further development
and extension of the approach \textit{per se} are of great interest.
To this end, several directions are seen. The most obvious modification
of the suggested approach is to apply it to problems opposite to the
one considered herein. Particularly, in many applications the task
is to change the behavior of the system by an external action \cite{Cornelius2013}.
From the nonlinear dynamics prospective, this means pushing the system
from one attractor to the basin of another. In this situation, quantification
of the stability threshold immediately provides the information about
the optimal perturbation to induce the switching.

Another possibility is to extend the concept of the stability threshold
to multistable systems and reversible transitions between different
attractors. In many applications, not a single, but several different
stable regimes play essential roles, and the system may switch these
regimes. The examples are switching between various patterns of activity
in neural circuits \cite{Klinshov2014,Hopfield82}, and episodic
emergence of El Ni\~no events \cite{Ludescher13}, transitions between
free flow and congestion in traffic \cite{Li15} or between failure
states and active states in self-recovering networks \cite{Majdandzic13}.
In all these cases, transitions in both directions are of interest.
To study such transitions from one attractor to another and backwards,
two stability thresholds can be introduced associated with each of
the transitions. The values of these thresholds may provide important
information about the transition rates when the system is externally
perturbed.

Finally, the numerical algorithm can be further developed and extended
for a broader class of systems. In particular, the current version
of the algorithm relies on smoothness of attraction basins borders,
which is a limitation of the method. It is known that basin boundaries
for some dynamical systems can be fractal \cite{Grebogi83}. In this
case the border can not be locally approximated by a tangential hyperplane,
which is crucial to determine the direction in which to move along
it. However, it is still possible to move along the border if the
steps are taken in random direction. Each step is accepted if it brings
the point closer to the attractor and declined otherwise. This modification
of the method requires additional investigation to the define the
conditions of its applicability and estimate its convergence.

\section*{Acknowledgments}
This paper was developed within the scope of the IRTG 1740/TRP 2011/50151-0, funded by the DFG/FAPESP, and supported by the Government of the Russian Federation (Agreement No. 14.Z50.31.0033 with the Institute of Applied Physics RAS). The first author thanks Dr. Roman Ovsyannikov for valuable discussions regarding estimation of the mistake probability.

\end{document}